\documentclass[12pt]{article} 

\usepackage{endnotes}

\let\footnote=\endnote

\usepackage{graphicx} 
\makeatletter
\renewcommand\@biblabel[1]{\textbf{#1.}} 
\renewcommand{\@listI}{\itemsep=0pt} 

\renewcommand{\maketitle}{ 
\begin{flushright} 
{\Large\@title} 

\vspace{50pt} 

{\large\@author} 
\\\@date 

\vspace{40pt} 
\end{flushright}
}

\title{\textbf{The ultimate tactics of self-referential systems\footnote{This essay received the 4th. Prize in the 2015 FQXi essay contest: ``Trick or Truth: the Mysterious Connection Between Physics and Mathematics".}}\\ 
 } 

\author{\textsc{Christine C. Dantas} 
\\ $~$\\ {\textit{Instituto de Aeron\'autica e Espa\c co \\AMR/IAE/DCTA - Brazil}\\ {\footnotesize {\tt christineccd@iae.cta.br, ccordulad@gmail.com}}\\$~$}} 

\date{\today} 

\begin{document}

\maketitle 



\begin{abstract} 
Mathematics is usually regarded as a kind of language. The essential behavior of physical phenomena can be expressed by mathematical laws, providing descriptions and predictions. In the present essay I argue that, although mathematics can be seen, in a first approach, as a language, it goes beyond this concept. I conjecture that mathematics presents two extreme features, denoted here by {\sl irreducibility} and {\sl insaturation}, representing delimiters for self-referentiality. These features are then related to physical laws by realizing that nature is a self-referential system obeying bounds similar to those respected by mathematics. Self-referential systems can only be autonomous entities by a kind of metabolism that provides and sustains such an autonomy. A rational mind, able of consciousness, is a manifestation of the self-referentiality of the Universe. Hence mathematics is here proposed to go beyond language by actually representing the most fundamental existence condition for self-referentiality.  This idea is synthesized in the form of a principle, namely, that {\sl mathematics is the ultimate tactics of self-referential systems to mimic themselves}. That is, well beyond an effective language to express the physical world, mathematics uncovers a deep manifestation of the autonomous nature of the Universe, wherein the human brain is but an instance.
\end{abstract}


\vspace{30pt} 


\clearpage

{\footnotesize \raggedleft{}``To the question which is older, day or night,\\
he $[$Thales of Miletus$]$ replied: \\
{\sl --- Night is the older by one day}"\footnote{Thales of Miletus, 624 BC--546 BC, Biographical Encyclopedia of Universal History.}. \\} 

\vspace{1cm}

{\footnotesize \raggedleft{}``Either mathematics is too big for the human mind\\ or the human mind is more than a machine." \\ Kurt G\"odel\\} 

\vspace{1cm}

\noindent {\large{\sl Introduction}}
\vspace{0.1cm}
\hrule
\vspace{0.5cm}
\noindent {\bf Outline ---}

The literature on the connection between mathematics and physics is vast, and will not be reviewed here (a good start is given by Penrose \cite{penrose}). The present theme is about one of most profound and disturbing mysteries ever touched by the human mind, and it is obviously very hard to find a definite answer at our present state of knowledge. The best one can do about that gigantic question is to offer some perceptions, specially considering a short text such as this.

The outline for this essay is the following. I review the idea that mathematics is, essentially, some kind of language, and argue to what extent this position can be sustained as a fundamental one. Here, mathematics is seen as something more than a language. I introduce two conjectures for mathematics and identify how these conjectures apply to nature. Finally, I state how these conjectures serve as guide for a principle that amalgamates mathematics and physics from an unusual point of view. 

\vspace{0.3cm}

\noindent {\bf To what extent is mathematics a language? ---}

A language is a form of expression and communication, which can be realized by different means, but it also involves some kind of transformation between two domains. The study of the methods, development, and applications of mathematics, indicates that it operates as a language.

Internally, mathematics is based on some set of non-contradictory, fundamental notions (axioms), which are constructively enlarged into more sophisticated statements and relations (theorems), by the incremental use of logical operations. Externally, mathematics mediates the description of relations from the sensible domain of phenomena into the domain of logical associations, which are, by themselves, internally consistent.

Actually, for whatever purpose, mathematics seems to be ``overly effective", because it can be used to describe, to a certain extent, anything one desires\footnote{For occurrences of the Golden Ratio in the natural world, see, e.g., Ref. \cite{livio}. For a beautiful example of two different partial differential equations (PDEs) describing the same physical phenomenon, see Ref. \cite{Bona}. A limiting procedure that suggests an explanation for the wide applicability and universality of some integrable PDEs can be found in \cite{Cal90}.}. There is always a means, under varying degrees of limitations, in which any sensible object, process, or relation can be expressed by a formal, deductive system, such as mathematics. The ubiquity of mathematical representations and transformations is sometimes regarded a trivial ``fact of the world", because it just implies that mathematics is wide and internally consistent enough to embrace any condition or aspect. 
But such a language must necessarily be of a very special kind, considering its  ``unreasonable effectiveness" \cite{Wig} to describe physical phenomena.   Is there a ``natural selectiveness effect", based perhaps on the own rigorous nature of mathematics, able explain its universality of application? Why is nature so mathematical?

I believe that there is no ``natural" explanation for the ``unreasonable effectiveness" --- or even the ``overly effectiveness" --- of mathematics if we regard it purely as a language, that is, from a fundamental point of view. My justification is based on the following observations:

a) When mathematics is seen as a language, it is merely a ``receptacle for logically configured relations" that summarize the essence of a physical phenomenon. Although it is, in a first approach, a reasonable view, it does not explain where the rules and meanings of the logical relations come from. The only possible way we could justify it as an ``effective receptacle" would be in a {\sl normative sense}, that is, in a prescriptive way. {\sl But such a normative way does not offer any intrinsic or fundamental connection with the natural phenomena, except if imposed arbitrarily}. 

b) When a physical law is expressed mathematically, no evidence for the ``truth" regarding the law actually surfaces, but only a guarantee for the logical foundation of the outcomes resulting from the implemented assumptions. In fact, any application of mathematics (seen as a form of language) to physical problems just brings forth a {\sl relative valorization of their logical evidence, but not the evidence, per se, of their truth}.

I firmly believe that mathematics is more than a ``transformation machine"\footnote{Although very useful, mathematics would be ultimately limited to transform ``emptiness into emptiness" (see commentary by Fr\`echet in  {\sl The Mathematical Thought}, part IV of Ref. \cite{Cav}).}. I regard mathematics as a language to the extent that it indeed {\sl serves} as a language, but such a serviceability cannot be an admissible explanation for why nature is  mathematical. In order to address this question, I raise two conjectures, which will serve as a basis for looking at the problem from a different perspective.

\vspace{0.3cm}

\noindent {\large {\sl Two Conjectures for the Foundations of Mathematics}}
\vspace{0.1cm}
\hrule
\vspace{0.5cm}

The proposed conjectures are philosophical, a fact that could be unattractive to some readers. However, these ideas could eventually be expressed in a more concrete or formal way, so they should be regarded as preliminary for the purposes of the present essay. It is clearly very hard\footnote{I am tempted to assert that such kinds of proofs are actually impossible.} to develop an independent methodology to avoid the ironic situation of using mathematical principles themselves in order to explain mathematics. This is not, evidently, the purpose here, so we limit to qualitative statements, on a more  ``meta", abstract level.

{\sc I. IRREDUCIBILITY: Mathematics is irreducible to anything else that is not itself mathematically expressible.}

\noindent {\bf What it means ---} Mathematics cannot be primarily reduced to anything different from itself. Any attempt to characterize mathematics, methodically, will necessarily involve the use of mathematically defined cons\-tituents and procedures (representations), so that {\sl an ultimate regression that is not mathematically based is impossible}.

\noindent {\bf Conjecture's nickname ---} The miniature extremeness of mathematics.

\noindent {\bf Connection with nature ---} Nature also presents a sense of irreducibility, but this property acts differently from mathematics.  Nature cannot be ultimately reduced to anything different from itself, because it is one whole essence that embeds everything inside itself. But what is for mathematics an intrinsic, defining property, for nature it serves as a delimitation: {\sl mathematics is an expression in the realm of the possible, whereas nature is an expression of what actually is}.

\vspace{0.5cm}

{\sc II . INSATURATION: Mathematics cannot, as a whole, be constructed from a ``master impredicative".}

\noindent {\bf What it means ---} First, an ``impredicative" is any self-referential definition, and self-reference occurs ``when a sentence, idea or formula refers to itself"\footnote{See, e.g.: {\tt http://en.wikipedia.org/wiki/Impredicativity} \\and {\tt http://en.wikipedia.org/wiki/Self-reference}, respectively.}. Second, this conjecture relies on the idea that countably infinite, self-referen\-cing mathe\-matical formulations can always be exhaustively specified\footnote{That statement is assumed valid in terms of {\sl circular quantifications} only. According to Wikipedia, ``The vicious circle principle is a principle that was endorsed by many predicativist mathematicians in the early 20th century to prevent contradictions. The principle states that no object or property may be introduced by a definition that depends on that object or property itself." It was later realized that circularity in terms of {\sl quantification} does not lead to paradoxes, as such definitions do not  ``create sets or properties or objects, but rather just give one way of picking out the already existing entity from the collection of which it is a part" (c.f. {\tt http://en.wikipedia.org/wiki/Vicious\_circle\_principle}).}, but mathematics itself is not {\sl saturated}. This means that mathematics cannot, as a whole, be constructed from a single, all encompassing self-referential defi\-nition, i.e., from a ``master impredicative".

\noindent {\bf Conjecture's nickname ---} The  vertigo extremeness of mathematics\footnote{It is just like ``looking down the wholeness of oneself from a great height", but, then, with vertigo, one cannot {\sl actually} look.}.

\noindent {\bf Connection with nature ---} Nature also presents a sense of insaturation, but, again, in a delimiting way. How can nature establish itself as 
a single, all encompassing self-referential essence? Nature is an open system towards becoming, in this sense, it is insaturated against its full self-definition. It opens a kind of ``space for realizations to come" \cite{Bergson_CE}: if we attempt to fully characterize nature, all its past must be integrated; yet, the future, which is potentially a part of it, can never enter into its determination, {\sl except as a prediction, as an expectation, and never as a realization}\footnote{Here I do not desire to engage in a discussion of various philosophical positions on the nature of time or determinism. Notice that the conjectures do not strictly depend on those.}. Is there a ``reference" to which the universe is {\sl not} an open system? Any answer to this question must involve a  sense of freedom from a delimitation, {\sl but such a freedom cannot be self-referential}.

\clearpage

\noindent {\bf Further commentaries on the Conjectures ---}

The first conjecture was partially based on observations already raised by the 20th. cent. French philosopher of mathematics, Jean Caivaill\`es\footnote{Sadly, he had an unfortunate end, being executed by the Gestapo at the end of WWII. Details can be found in Ref. \cite{Cav}, or on Wikipedia, {\tt http://en.wikipedia.org/wiki/Jean\_Cavailles}.}. He wrote: ``{\sl Mathematics constitutes a becoming, that is, an irreducible reality to something else different from it}"  (my translation, from Ref. \cite{Cav}). Cavaill\`es' observation, as I interpret it, corroborates with the idea that there is no ``meta-language out there", nor ``within", which rigorously defines mathematics, {\sl non-mathematically}, to its most basic constituents\footnote{Indeed, it is quite difficult to characterize mathematics, {\sl at the level of rigor that it attains as its natural condition}, by some different means. For example, according to Cavaill\`es \cite{Cav}: (i) mathematics is not exactly a part of logic, since some mathematical notions may be combinatorial in nature, or based on other notions irreducible to purely logical operations --- yet, those are still mathematical anyway. And (ii) mathematics cannot be entirely characterized as a hypothetical-deductive system, yet alternative/complementary notions for that purpose are still mathematical anyway.}.  Mathematics allows innumerable different representations for the same given concept, relation, or object, and can also be constructed from other basic objects\footnote{Not necessarily sets \cite{Topoi},\cite{lane}}. Yet, any such representation or construction is still completely mathematical. Any precise view of mathematics is mathematical. It all proceeds as if mathematics locked within itself, in order to {\sl effectively realize what it is, as if nothing else could play that unique role}.

The second conjecture was partially based on an informal use of the G\"odel's theorems (c.f., e.g., Ref. \cite{Hof}), as well as on various considerations arising in computer science and mathematical logic. The rationale summarizing those considerations and united in the conjecture is the following. There is a sense in which mathematics is flexible enough to provide a system able to describe an autonomous exhaustion of all consistent logical possibilities within itself. However, it cannot be itself embedded into those same infinitely comprehensive possibilities, that is, {\sl it cannot be constructed from a single and consistent self-definition ``on top"}. Such an all-encompassing construction, that is, completely self-referential {\sl and} whole, cannot be ``seen" inside the mathematical realm, given whatever consistent set of axioms. Such a consistency can be achieved in principle, but  not proven within the system \cite{Hof}.  In a sense, a ``view from outside" is required. Hence, mathematics is not saturated, and {\sl specially not so from a ``meta point-of-view" of self-referentiality}. 

\noindent {\bf An important note on ``autonomous self-referential systems"---} I argue that the two conjectures together express  limiting conditions for {\sl autonomous self-referential systems}, and that these conditions are optimal, by construction. It is important to note here the reference to ``autonomy", in the present context, connected with the notion of self-referential systems. Two observations follow in this note, in order to clarify the principle stated afterwards. First, an autonomous self-referential system is irreducible to anything else that is not itself self-referential, a feature which express a kind of limit: the most elementary self-referential expression must be the primordial one (the generating ``seed"), otherwise the system cannot be autonomous, in the sense of self-generating. On the other hand, if the autonomous self-referential system is reducible to a predicative, then it is non-referential, which contradicts the initial assumption that it is. Second, an autonomous self-referential system must contain an assertion ``on top" that is not fundamentally self-referential, in order to close the ladder of all self-referential expressions, thus admitting an {\sl external reference to which it can be defined as autonomous}\footnote{We allow for the existence of infinitely countable predicatives within the system, as particular cases of impredicatives that are self-referring to a null set.}. 

\vspace{0.3cm}

\noindent {\large {\sl A Principle}}
\vspace{0.1cm}
\hrule
\vspace{0.3cm}

The two proposed conjectures reflect ``small" and ``large" scale features, in an abstract sense, and share a delimiting correspondence with the physical Universe.  Indeed, any ``reducible essence" of the Universe, whatever its constitution, is always a part of the Universe and cannot be reduced to any ``different reality". In the same manner, there is no possible way for the Universe to ``become the largest essence of all",  under its own unfolding state. The following principle proposes a view for those considerations by stating that mathematics is a {\sl condition}, whereas the Universe is a {\sl manifestation}, and both are inseparable when approached from the point of view of a conscious mind.

\vspace{0.5cm}

{\sc PRINCIPLE: Mathematics is an existence condition for autonomous self-referential systems, in particular, the Universe.} 

\noindent {\bf What it means ---}  This principle provides a synthesis of both conjectures in the sense that {\sl mathematics uncovers the autonomous self-referentiality of the Universe}, providing, more than a language, a metabolism between conscience and nature\footnote{Far from bringing ideological assumptions to the matter, compare  with Marx's quote: {\it ``Labour (...)  is an eternal natural necessity which mediates the metabolism between man and nature, and therefore human life itself"} \cite{marx}. This made me think of a parallel reasoning for the connection of  mathematics and nature.}. In a stronger sense, {\sl mathematics should arise ``with" (note: not ``in") any autonomous self-referential universe}, that is,  any universe with conditions for the emergence of conscious beings. 

\noindent {\bf Principle's nickname ---} Mathematics represents the ultimate {\sl tactics} of self-referential systems to mimic themselves.

\noindent {\bf Explanation ---} Mathematics does not seem to depend on nature, because there is an indefinitely large number of mathematical constructs and theorems that do not have an immediate application at all, but yet they are eventually discovered and proved independently of the physical world.  Many of such mathematical constructions, initially disembodied from a physical application, do end up as being excellent descriptions of new phenomena as soon as we, as humans, construct the necessary relations through observation and theoretical reasoning. Indeed, the human brain is but an instance of the self-referentiality of the Universe (we are just atoms thinking about atoms \cite{sagan}). This indicates a firm evidence for the {\sl mathematical knowability of self-referential systems}. This {\sl knowability} is the act of the system to reproduce whatever it is part of, hence, to {\sl mimic itself}$~$\footnote{Indeed, one may argue that a sufficiently advanced technology must be  indistinguishable from nature. That would be an expected trend, according to the arguments of the present essay.\\ {\tt http://www.universetoday.com/93449/do-alien-civilizations-inevitably-go-green/}}. In the present view, mathematics acts as a ``conditioner" of all logical possibilites. It is seen as a language only {\sl within the act of making assertions about what is manifested} in the physical world. I believe that this {\sl knowability} indicates that the ``reality of mathematics" arises {\sl with} the ``reality of the Universe", as a single act.  This act is completely crystallized, waiting to be discovered. This {\sl self-intelligibility} is a manifestation of the Universe, {\sl within ourselves}. Mathematics is our tactics to navigate in the ocean that can be known. 

\clearpage

\noindent {\large {\sl Final Reflections and Self-Reflections}}
\vspace{0.1cm}
\hrule
\vspace{0.5cm}

The only way for a self-referential system to autonomously exist is to have {\sl bounds} that provide such an autonomy, bounds that I have identified in extreme features of mathematics, delineated in the two conjectures.  As stated previously, mathematics is the realm of the possible (condition); nature is the realm of what actually is (manifestation). {\sl The  ultimate ``tactics" of knowability arises from exhausting that same knowability}. Mathematics and physical laws are, as I attempted to show, a pure self-metabolism, seeking and longing for {\sl that} explanation of itself ({\sl and what shall that explanation be?}). 

According to Galileo Galilei,
\begin{quote}
Philosophy is written in that great book which ever lies before our eyes ---  I mean the universe --- but we cannot understand it if we do not first learn the language and grasp the symbols, in which it is written. This book is written in the mathematical language, and the symbols are triangles, circles and other geometrical figures, without whose help it is impossible to comprehend a single word of it; without which one wanders in vain through a dark labyrinth\footnote{{\tt http://en.wikiquote.org/wiki/Galileo\_Galilei}.}.
\end{quote}

Indeed, we first learn the language. But, going beyond the language, we turn around to ourselves and see that {\sl we, nature, are the book}, written inside a book, inside another book... A dark labyrinth gives its place to a beautiful spiral. Or a dot. Or a number. Or the infinite.  The ``effectiveness of mathematics in the natural sciences" is unreasonable and will always be.  That is our tactics to keep on turning the pages.

\vspace{0.5cm}
\hrule 
\vspace{0.1cm}
\hrule

\vspace{1cm}

{\bf Acknowledgements.} The author thanks Fabiano L. de Sousa for useful suggestions.

\clearpage

     \parskip 0.6ex

\theendnotes

\clearpage 

\bibliographystyle{unsrt}

\bibliography{References}


\end{document}